\documentclass[english]{emulateapj}

\usepackage[T1]{fontenc}
\usepackage[latin9]{inputenc}
\usepackage{array}
\usepackage{float}	
\usepackage{amsmath}
\usepackage{color}

\makeatletter

\providecommand{\tabularnewline}{\\}

\@ifundefined{textcolor}{}
{%
 \definecolor{BLACK}{gray}{0}
 \definecolor{WHITE}{gray}{1}
 \definecolor{RED}{rgb}{1,0,0}
 \definecolor{GREEN}{rgb}{0,1,0}
 \definecolor{BLUE}{rgb}{0,0,1}
 \definecolor{CYAN}{cmyk}{1,0,0,0}
 \definecolor{MAGENTA}{cmyk}{0,1,0,0}
 \definecolor{YELLOW}{cmyk}{0,0,1,0}
 }

\makeatother

\usepackage[normalem]{ulem}  

\usepackage{babel}

\usepackage{natbib}

\shorttitle{Upper bounds on {\it r}-mode amplitude in LMXB NSs}
\shortauthors{Mahmoodifar \& Strohmayer}

\begin{document}

\title{Upper bounds on {\it\lowercase{r}}-mode amplitudes from observations of low-mass X-ray binary
neutron stars}

\author{Simin Mahmoodifar$^1$ and Tod Strohmayer$^2$ \\ {\normalfont $^1$Department of Physics, University of Maryland College
Park, MD 20742, USA} \\ {\normalfont $^2$Astrophysics Science Division, NASA's
Goddard Space Flight Center, Greenbelt, MD 20771, USA}} 

\begin{abstract}

We present upper limits on
the amplitude of {\it r}-mode oscillations, and
gravitational-radiation-induced spin-down rates, in low mass X-ray
binary neutron stars under the assumption that the quiescent
neutron star luminosity is powered by dissipation from a steady-state
{\it r}-mode. For masses $<
2 M_{\odot}$ we find dimensionless {\it r}-mode amplitudes in the range
from about $1\times 10^{-8}$ to $1.5\times 10^{-6}$. For the accreting
millisecond X-ray pulsar sources with known quiescent
spin-down rates these limits suggest that $\lesssim 1\%$ of the
observed rate can be due to an unstable {\it r}-mode.  Interestingly, the
source with the highest amplitude limit, NGC 6440,
could have an {\it r}-mode spin-down rate comparable to the observed,
quiescent rate for SAX J1808-3658. Thus, quiescent spin-down
measurements for this source would be particularly interesting. For
all sources considered here our amplitude limits suggest
that gravitational wave signals are likely too weak for
detection with Advanced LIGO.  Our highest mass model ($2.21\,
M_{\odot}$) can support enhanced, direct Urca neutrino emission in the
core and thus can have higher {\it r}-mode amplitudes.  Indeed, the inferred
{\it r}-mode spin-down rates at these higher amplitudes are inconsistent
with the observed spin-down rates for some of the sources, such as IGR
J00291+5934 and XTE J1751-305. In the absence of other significant
sources of internal heat, these results could be used to place an
upper limit on the masses of these sources if they were made of hadronic matter, or alternatively it could be used to probe the
existence of exotic matter in them if their masses were known.


\end{abstract}
\keywords{dense matter -- gravitational waves -- stars: neutron -- 
stars: oscillations -- stars: rotation -- X-rays: binaries}

\section{Introduction}

The {\it r}-modes are large-scale currents in neutron stars (NSs) that couple to
gravitational radiation and remove energy and angular momentum from
the star in the form of gravitational waves
\citep{Andersson:1997xt,Friedman:1997uh,Friedman:1978hf,Lindblom:1998wf}. The
physics of these oscillations is important in relating the microscopic
properties of dense matter--such as its viscosity and neutrino
emissivity--to the macroscopic, observable properties of NSs
such as their spin frequency, temperature, mass and radius. The
{\it r}-modes are unstable to gravitational radiation and their amplitudes
can grow exponentially if viscous and other possible damping
mechanisms are not large enough. Because they can affect the dynamic
properties of the star--its spin frequency and temperature
evolution--they are potentially important probes of the phases of
dense matter. For example, the boundary of the {\it r}-mode instability
region in the spin frequency - temperature plane, and in particular
its minimum, which may determine the final rotation frequency of the
star, is very different for stars with different interior compositions
\citep{Alford:2010fd,Ho:2011tt,2012MNRAS.424...93H}.

Whether or not the {\it r}-mode instability limits the spin rates of some
NSs it is nevertheless important to explore its
potential astrophysically observable signatures.  While mass-radius
measurements of NSs are important for constraining the
equation of state (EOS) of dense matter, observations of their dynamic
properties such as spin and thermal evolution are important and
potentially more efficient in discriminating between different phases
of dense matter. That is because dynamic properties are affected by the
transport and thermodynamic properties of dense matter inside the
star, such as viscosity, heat conductivity and neutrino
emissivity which depend on low energy degrees of freedom and are very
different depending on the phase of dense matter present
\citep{Alford:2010fd,Alford:2012jc,Alford:2012yn}.

Pulse timing observations of accreting millisecond X-ray pulsars
(AMXPs) made with NASA's {\it Rossi X-ray Timing Explorer} ({\it RXTE}) are
beginning to reveal the long-term spin evolution of low mass X-ray
binary (LMXB) NSs.  The pulsar recycling hypothesis whereby
millisecond radio pulsars acquire their fast spins via accretion,
requires that at least some of these stars are spun-up to hundreds of
Hz--the current NS spin record being 716 Hz
\citep{2006Sci...311.1901H}.  Still, it remains somewhat puzzling that
the spin frequency distribution of AMXPs appears to cut-off well below
the mass-shedding limit of essentially all realistic NS EOS
\citep{2003Natur.424...42C}, and it has been suggested that the {\it r}-mode
instability may play a role in limiting NS spin rates
\citep{Andersson:1998qs}. Alternatively, recent work supports the idea
that at least some of the AMXP population are in or close to ``spin
equilibrium,'' essentially determined by the physics of magnetized
accretion, and that an additional mechanism, such as the {\it r}-mode
torque, is not required to halt their spin-up
\citep{1997ApJ...490L..87W,2012ApJ...746....9P}.

While pulse timing noise in AMXPs has made the interpretation of the X-ray
pulsation data difficult, there are now several convincing
measurements of spin-down during the quiescent phases between
accretion outbursts. For example, both SAX J1808.4-3658 (hereafter SAX
J1808) and IGR J00291+5934 (hereafter IGR J00291) show spin-down of
the NS between outbursts, at rates, $\dot\nu$, of
approximately $1-3 \times 10^{-15}$ Hz s$^{-1}$
\citep{Patruno:2010qz,2009ApJ...702.1673H}.  The most convincing
evidence for an accretion-induced spin-up during outbursts is in IGR
J00291, for which a peak value of about $3 \times 10^{-13}$ Hz
s$^{-1}$ has now been inferred \citep{Patruno:2010qz}.  The magnitude
of the ``instantaneous'' spin-up torque is larger in this source than
the spin-down torque by about a factor of 100.  The long-term
evolution is then a competition between the accretion outbursts that
spin it up and the quiescent spin-down intervals.  If the outbursts
are frequent enough then the star will be spun-up, if not, then
long-term spin-down occurs. As the spin-up torque is proportional to
the mass accretion rate, the outcome can also be expressed in terms of
the long-term, average mass accretion rate.

The observed quiescent spin-down rate, $\dot{\nu}_{sd}$, puts an upper
limit on the torque that can be present due to any {\it r}-mode, as it has
to be less than the observed torque of $2\pi I \dot\nu_{sd}$, where
$I$ is the moment of inertia of the NS.  Now, spin-down of
pulsars is typically ascribed to the magnetic-dipole torque that is
almost certainly present at some level due to the large scale magnetic
field of the NS.  Indeed, magnetic field estimates are
usually obtained by equating the {\it entire} observed spin-down to
that expected theoretically for magnetic-dipole radiation, but the
observed spin-down is due to the sum total of any torques present.
Assuming that both {\it r}-mode and magnetic-dipole torques are present,
then the magnitude of the {\it r}-mode torque is equal to the observed
torque minus the magnetic-dipole torque.  Since the magnetic
field strengths of AMXPs are typically not known independently from
spin-down estimates, the precise value of the magnetic-dipole torque
is not known a priori.  However, to the extent that the magnetic
torque accounts for the majority of the observed spin-down, as is
typically assumed, then the {\it r}-mode torque is likely to be much less
than the measured spin-down torque.

In addition to torquing the NS, the viscous damping of the
{\it r}-modes acts as an internal source of heat. Calculations of the
coupled thermal and spin evolution of accreting NSs
including the effects of {\it r}-mode heating and gravitational radiation
have been carried out by several authors \citep{2000ApJ...536..915B,
2011ApJ...738L..14H, Ho:2011tt, 2012MNRAS.424...93H}.  The primary
assumption in these analyses has been that the long-term accretion
(spin-up) torque is balanced by the {\it r}-mode torque due to the emission
of gravitational radiation that carries away angular momentum, that
is, the long-term average accretion torque is in equilibrium with the
{\it r}-mode torque. As noted above, the recent pulse timing observations of
AMXPs indicate that this equilibrium assumption is likely not realized
in practice. Moreover, \citet{2000ApJ...536..915B} showed that if the
average, fiducial accretion torque given by $N_{acc} = \langle \dot M
\rangle (G M R)^{1/2}$, where $\langle \dot M \rangle$ is the
long-term average mass accretion rate, were balanced by the {\it r}-mode
torque then the quiescent luminosities of some accreting NS
transients should be substantially larger than observed due to heating
from the {\it r}-modes.  Here we use a similar argument to place upper
bounds on the {\it r}-mode amplitudes that can be present in accreting
NSs assuming that {\it r}-mode heating provides the source of
NS luminosity in the absence of accretion.  We then use
these amplitude limits to assess the level of {\it r}-mode spin-down that
can be present, its relation to observed spin-down rates when
available, and the expected strength of gravitational radiation.

The paper is organized as follows. In Section 2 we review the basic theory
of the {\it r}-mode instability and how it couples to the thermal and spin
evolution of NSs. In Section 3 we describe our methods for
constraining the {\it r}-mode amplitudes using the observed properties of
quiescent, LMXB NSs, and we assess the implications for
{\it r}-mode spin-down and the emission of gravitational radiation. We
provide a summary and discussion of the implications of our findings
in Section 4.

\section{Spin-down due to unstable {\it\lowercase{r}}-modes}
The ``flow pattern'' of the {\it r}-modes is prograde in the inertial frame
and retrograde in the rotating frame, which means that it moves in the
same direction as the star's rotation as seen by an observer at
infinity, but in the opposite direction as seen by an observer at rest
on the star. Any mode that is retrograde in the co-rotating frame and
prograde in the inertial frame grows as a result of its emitting
gravitational waves. This is the well-known
Chandrasekhar-Friedman-Schutz mechanism, and it means that
gravitational radiation drives the mode rather than dampening
it. Viscosity on the other hand tends to damp the {\it r}-mode by
transferring angular momentum from the mode to the rigid (unperturbed)
star. The total angular momentum of a perturbed star can be written
as
\begin{equation}
J=I\Omega- J_c 
\end{equation}

where $I$ is the moment of inertia and $\Omega$ is the angular
rotation frequency of the star and $J_c$ is the canonical angular
momentum of the mode, given by
\begin{equation}
J_c=-\frac{3}{2} \tilde{J}  M R^2 \Omega \alpha^2\label{eq:J_c}
\end{equation}

where $\alpha$ is the amplitude of the {\it r}-mode and $\tilde{J}$ is a
dimensionless constant defined by \citep{Owen:1998xg}
\begin{equation}
\tilde{J} \equiv \frac{1}{M R^4}\int_0^R \rho r^6 dr\label{eq:J-tilde}
\end{equation}
where $\rho$ is the run of density within the neutron
star.  The moment of inertia of the star, $I$, can also be written as
$I=\tilde{I} M R^2$ where $\tilde {I}$ is defined by
\begin{equation}
\tilde{I} \equiv \frac{8 \pi}{3 M R^2}\int_0^R \rho r^4 dr \; .
\end{equation}

To derive the equations for the dynamical evolution of the star, we use
the following argument by \citet{Ho:1999fh}.
The canonical angular momentum of the mode increases through
gravitational radiation and decreases by transferring angular momentum
to the star through viscosity

\begin{equation}
\frac{dJ_c}{dt}=-\frac{2}{\tau_{G}}J_c-\frac{2}{\tau_V}J_c\label{eq:dJ_c}
\end{equation}

where the viscous damping time $\tau_V$ is given by
$\frac{1}{\tau_V}=\frac{1}{\tau_S} + \frac{1}{\tau_B}+ ...$.  Here,
$\tau_S$ and $\tau_B$ refer to shear and bulk viscosities,
respectively, and the ellipsis denotes other possible dissipative
mechanisms, such as boundary layer effects
\citep{Wu:2000qy,Bildsten:2000ApJ...529L..33B}.

The second evolution equation is obtained by writing the conservation
of the total angular momentum $J$ of the perturbed star which says
that the total angular momentum of the star decreases due to
gravitational radiation and increases due to accretion

\begin{equation}
\frac{dJ}{dt}=-\frac{2J_c}{\tau_G}+N_{acc}\label{eq:dOmega} \;,
\end{equation}

where $N_{acc}$ is the accretion torque. For a fiducial torque
$N_{acc}$ can be written as $\dot{M}(G M R)^{1/2}$ which assumes that
each accreted particle transfers to the star an angular momentum equal
to the Keplerian value at the stellar radius $R$
\citep{2000ApJ...536..915B}. In the previous equations the quantities
$\frac{1}{\tau_i}$, where $i$ is either $G$, $B$, or $S$ for
gravitational radiation, bulk viscosity or shear viscosity timescales,
respectively, are given by,

\begin{equation}
\frac{1}{\tau_i}=-\frac{P_i}{2E_c} 
\end{equation}

where $E_c$ is the canonical energy of the {\it r}-mode, $P_G$ is the power
radiated by gravitational waves and $P_B$ and $P_S$ are the powers
dissipated due to bulk and shear viscosity, respectively (in natural
units where $c=\hbar=k_B=1$) \citep{Owen:1998xg,Alford:2010fd},
\begin{align}
E_c &=\frac{1}{2} \alpha^2 \Omega^2 \tilde{J} M R^2\\
P_G &=\frac{32\pi(m-1)^{2m}(m+2)^{2m+2}}{((2m+1)!!)^2(m+1)^{2m+2}}\tilde{J}_m^2GM^2R^{2m+2}\alpha^2\Omega^{2m+4}\\
P_B &=-\frac{16m}{(2m+3)(m+1)^5\kappa^2}\frac{\tilde{V}_m\Lambda_{{\rm QCD}}^{9-\delta}R^7\alpha^2\Omega^4T^\delta}{\Lambda_{EW}^4}\\
P_S &=-\frac{(m-1)(2m+1)\tilde{S}_m\Lambda_{{\rm QCD}}^{3+\sigma}R^3\alpha^2\Omega^2}{T^\sigma} \; . \end{align}

\begin{table*}
\renewcommand{\arraystretch}{1.5}
\caption{Parameters of the Neutron Star Models\label{tab:viscosity-parameters}}
\scalebox{0.95}{
\begin{tabular}{crrrrrrrrrrrrrr}
\tableline\tableline
Neutron Star & Shell & $R$ (km) &$\Omega_K$ (Hz) &$\tilde{I}$& $\tilde{J}$ & $\tilde{S}$ & $\tilde{V}$&$\tilde{C}_V$ & $\tilde{L}$&$\sigma$&$\delta$&$v$&$\theta$\\
\tableline
NS $1.4\, M_{\odot}$ & Core & $11.5$&$6020$&$0.283$ & $1.81\times10^{-2}$ & $7.68\times10^{-5}$ & $1.31\times10^{-3}$&$2.36\times10^{-2}$ & $1.91\times10^{-2}$&$\frac{5}{3}$&$6$ &$1$&$8$ \\
NS $2.0\, M_{\odot}$ &Core  & $11.0$&$7670$& $0.300$&$2.05\times10^{-2}$ &$2.25\times10^{-4}$ & $1.16\times10^{-3}$&$2.64\times10^{-2}$ & $1.69\times10^{-2}$&$''$&$''$ &$''$&$''$\\
NS $2.21\, M_{\odot}$ & m.U. core &$10.0$&$9310$&$0.295$& $2.02\times10^{-2}$ & $5.05\times10^{-4}$ & $9.34\times10^{-4}$&$2.62\times10^{-2}$ & $1.29\times10^{-2}$&$''$&$''$ &$''$&$''$\\
 & d.U. core && & & & &$1.16\times10^{-8}$& & $2.31\times10^{-5}$&&$4$ &&$6$ \\
\tableline
\end{tabular}}
\tablecomments{Radius, Kepler frequency and radial integral parameters
that appear in the moment of inertia, angular momentum of the mode,
dissipative powers due to shear viscosity (from leptonic interactions)
and bulk viscosity (due to Urca processes), specific heat and neutrino
luminosity for different NS models considered in this work
\citep{Alford:2012yn,Alford:2010fd}. For the $2.21\, M_{\odot}$ model
the bulk viscosity and neutrino luminosity parameters are different in
the inner core where direct Urca processes are allowed, therefore
these values are given separately in the last row.}
\end{table*}

Here we consider $m=2$ {\it r}-modes, and $\Lambda_{{\rm QCD}}$ and
$\Lambda_{{\rm EW}}$ are characteristic strong and electroweak scales
introduced to make $\tilde{V}$ and $\tilde{S}$ dimensionless. In our
calculations we have used $\Lambda_{QCD}=1$ GeV and $\Lambda_{EW}=100$
GeV.  The dimensionless parameters $\tilde{V}$, $\tilde{S}$,
$\tilde{I}$ and $\tilde{J}$, which involve radial integration over the
star, and $\delta$ and $\sigma$ are given in
Table~\ref{tab:viscosity-parameters} for three different NS
models that we study in this paper. All three models here are made of
non-superfluid, hadronic {\it npe} matter with the APR EOS
\citep{Akmal:1998cf}, which generates a reasonable NS mass-radius
relation that is consistent with observational constraints
(labeled AP4 in \citet{Hebeler:2013nza} and
\citet{Lattimer:2012nd}), but they have different masses ($1.4\,
M_{\odot}$, $2.0\, M_{\odot}$ and $2.21\, M_{\odot}$) and radii
\citep{Alford:2010fd}. The two models with masses of $1.4\, M_{\odot}$
and $2.0\, M_{\odot}$ only allow modified Urca neutrino emission in
the core, but the one with a mass of $2.21\, M_{\odot}$ allows direct
Urca neutrino emission in a core of radius $5.9$ km. Direct Urca
processes are very sensitive to the proton fraction of dense
matter. The required proton fraction is roughly $14 \%$ in the case of
the APR EOS, reached at relatively high density $n \sim 5n_0$, where
$n_0$ is the nuclear saturation density, this could be different for
other EOSs \citep{Alford:2010fd}.

The evolution equations for the amplitude of the {\it r}-mode, $\alpha$, and
spin frequency of the star, $\Omega$, can be written by using
Equations \ref{eq:dJ_c} and \ref{eq:dOmega} and substituting $J_c$
from Equation~\ref{eq:J_c}. The third equation --which describes the
temperature evolution-- can be obtained by noting
that the temperature of the star decreases due to thermal emission
from the surface and neutrino emission from the interior, which in an
average mass hadronic star is dominated by modified Urca processes (in
a massive star direct Urca processes can also occur in the core and
should be included in the neutrino emissivity as well), and it
increases due to the viscous dissipation of the {\it r}-mode energy, $P_V$.
This gives the following equations for the evolution of spin
frequency, {\it r}-mode amplitude and temperature,

\begin{subequations}
\begin{align}
\frac{d\Omega}{dt}&=-2Q\frac{\Omega\alpha^2}{\tau_V}+\frac{N_{acc}}{I}\label{eq:evolution-Omega}\\
\frac{d\alpha}{dt}&=-\frac{\alpha}{\tau_G}-\frac{\alpha}{\tau_{V}}(1-\alpha^2Q)-\frac{\alpha}{2 \Omega}\frac{N_{acc}}{I}\label{eq:evolution-alpha}\\
\frac{dE}{dt}&=C_V\frac{dT}{dt}=-L_{\nu}-L_{\gamma}+|P_V|+H \; , \label{eq:evolution-T}
\end{align}\label{eq:evolution}
\end{subequations}

where $Q\equiv\frac{3\tilde{J}}{2\tilde{I}}$, and the viscous
dissipated power is $P_V = P_S + P_B + \ldots$, and again the ellipsis
denotes other possible dissipative processes. In
Equation~(\ref{eq:evolution-T}), $H$ represents other heating mechanisms that
might be present in the star, but are not related to the {\it r}-mode
dissipation, such as deep crustal heating due to nuclear reactions in
the NS crust \citep{2003A&A...404L..33H, 2007ApJ...662.1188G}. Since
our goal in this paper is to obtain {\it upper} limits on {\it r}-mode
amplitudes we can safely set $H=0$. We discuss this further in the
next section.  Here, $L_{\nu}$, $L_{\gamma}$ and $C_V$, which are the
total neutrino luminosity, photon luminosity and specific heat of the
star, are given by (in natural units)

\begin{align}
L_{\nu} &=\frac{4\pi R^3\Lambda_{QCD}^{9-\theta}\tilde{L}}{\Lambda_{EW}^4}T^{\theta}\label{eq:neutrino-luminosity}\\
L_{\gamma} &=4\pi R^2 \sigma T_{eff}^4\label{eq:thermal-luminosity}\\
C_V &=4\pi\Lambda_{QCD}^{3-v}R^3\tilde{C}_VT^v \; ,\end{align}

where $T$ is the core temperature of the star and $T_{eff}$ is the
surface temperature. The dimensionless parameters $\tilde{L}_{\nu}$
and $\tilde{C}_V$ (that involve radial integration over the star \citep{Alford:2012yn}) and
$\theta$ and $v$ are given in Table~\ref{tab:viscosity-parameters} for
different stellar models.

\begin{figure}
%
\includegraphics[scale=0.9]{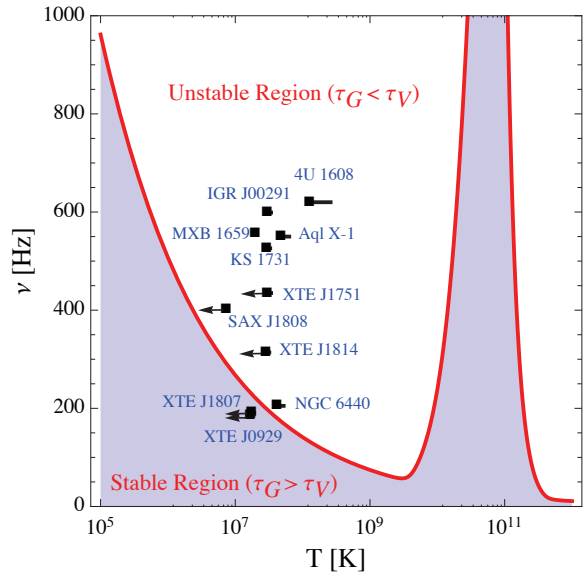}%
\caption{\label{fig:Omega-T} The {\it r}-mode instability region for a $1.4\,
M_{\odot}$ NS constructed with the APR EOS
in the spin frequency vs. core temperature plane. Also shown are some of 
the LMXBs which have been considered in 
this work. The horizontal line extending rightward from the temperature 
symbols (the black squares) shows the difference between two models for 
relating the surface temperature to the core temperature (i.e. the difference
between a fully or partially accreted envelope). The difference
between the core temperatures in these two cases gets larger as the
surface temperature increases, but even for the highest surface
temperature considered in this work the difference is not large enough
to change our results.\\
(A color version of this figure is available in the online journal.)}
\end{figure}

Figure~\ref{fig:Omega-T} shows the {\it r}-mode instability window computed
for a $1.4\, M_{\odot}$ NS using the APR EOS (the
modification to the instability window is rather modest in the case of
the two other stellar models considered in this paper
\citep{Alford:2010fd}). In the region above the curve the
gravitational radiation timescale is smaller than the viscous damping
timescales, therefore {\it r}-modes are unstable and their amplitudes grow
exponentially in this region. As can be seen in this figure most of
the LMXB's are in the unstable region for normal hadronic stars where
the damping is due to shear viscosity from leptonic scattering
(i.e. electron-electron and electron-proton scatterings, which are the
dominant contributions to the shear viscosity of normal hadronic
matter in NS cores) and bulk viscosity due to Urca processes,
therefore, there must be some non-linear mechanism that saturates the
amplitude of the unstable {\it r}-modes at a finite value. Supra-thermal
bulk viscosity \citep{Alford:2010gw} is one of these non-linear
mechanisms, but in the case that only the core of the star is
considered and the effects of the crust are ignored, it can only
saturate the {\it r}-mode at large amplitudes, ($\alpha \sim 1$)
\citep{Alford:2011pi}. Magnetohydrodynamic coupling to the stellar
magnetic field is another mechanism that can damp the {\it r}-mode
instability, but it can only saturate the {\it r}-mode at large amplitudes
($\alpha \gtrsim 0.01$) in the presence of a magnetic field
significantly larger than $\sim 10^{8}$ G that is characteristic of
the LMXB sources considered here
\citep{2000ApJ...531L.139R,2001PhRvD..64j4014R}. Mode coupling can
saturate the {\it r}-mode at smaller amplitudes ($\alpha \sim 10^{-4}$), but
those values are still very large compared to the upper limits we find
in this work
\citep{Arras:2002dw,2007PhRvD..76f4019B,2009PhRvD..79j4003B}. At this
point it is not entirely clear which mechanism is actually responsible
for saturating the {\it r}-mode amplitude (none of the saturation mechanisms
proposed so far can saturate {\it r}-modes at the low amplitudes we find
here), however, in this paper our primary interest is to obtain upper
bounds on {\it r}-mode amplitudes from observations of NS
transients and this does not require a precise understanding of the
saturation mechanism. Understanding the detailed physics of the
saturation mechanism is an important issue, but it is beyond the scope
of this paper.

Here we note that if we were to consider the existence of exotic forms
of matter, such as strange quark matter in the star, then the shape of
the instability window would change due to their different shear and
bulk viscosities, and it is possible that the LMXBs considered here
might fall outside of the resulting instability window
\citep{Alford:2010fd,2012MNRAS.424...93H,Schwenzer:2012ga}. The shape
of the instability window may also be different due to crust boundary
layer effects \citep{Ho:2011tt}, but even in that case for realistic
boundary layer models most of the LMXB sources are in the unstable
region. Therefore, we assume that these sources have unstable {\it r}-modes
which are emitting gravitational waves but the amplitude of the {\it r}-mode
is not growing exponentially, and there is a mechanism that can
saturate the growth of the {\it r}-mode. In addition, we assume that all of
these stars are made of normal hadronic matter (constructed with the
APR EOS) and all of the sources that are in the unstable region for
hadronic stars in Figure~\ref{fig:Omega-T} are emitting gravitational
waves at a constant amplitude.

Since we do not know which mechanism is actually responsible for
saturating the {\it r}-mode amplitude, we simply assume that there is a
nonlinear mechanism that saturates the mode. When $\alpha$ hits the
saturation amplitude, the right hand side of
Equation~(\ref{eq:evolution-alpha}) becomes zero which implies that
\begin{align}
\frac{1}{\tau_V} &=\frac{1}{\tau_G}\frac{1}{1-\alpha^2 Q} \; ,
\end{align}
where we have neglected the last term in Equation~(\ref{eq:evolution-alpha})
since it is much smaller than the other terms.

Therefore, when the amplitude is saturated, in all of the evolution
equations $\frac{1}{\tau_V}$ should be replaced by
$\frac{1}{\tau_G}(\frac{1}{1-\alpha^2 Q})$ and $|P_V|$ by
$\frac{P_G}{1-\alpha^2 Q}$. Since viscosity alone cannot stop the
growth of the {\it r}-mode and an extra mechanism is needed to do that,
therefore at saturation $\frac{1}{\tau_G}(\frac{1}{1-\alpha^2 Q})$ is
larger than $\frac{1}{\tau_V}$ and the reheating term on the right
hand side of the temperature evolution equation
(Eq~\ref{eq:evolution-T}) will be larger than the reheating due to
bulk and shear viscosity \citep{Alford:2012jc}.

\begin{table}
\renewcommand{\arraystretch}{1.5}
\caption{Constraints from Spin Equilibrium\label{tab:spin-equilibrium alpha}}
\scalebox{0.75}{
\begin{tabular}{crrrrr}
\tableline\tableline
Source & $\Delta=\frac{t_o}{t_r}$&  $\dot{\nu}_{su}$& $\alpha_{sp.eq} $ &  $\alpha_{sp.eq} $ &  $\alpha_{sp.eq} $ \tabularnewline
&& (Hz s$^{-1}$) &$(1.4\, M_{\odot})$& $(2.0\, M_{\odot})$ & $ (2.21\, M_{\odot})$\tabularnewline
\tableline
IGR J$00291$&$\frac{13}{1363} $& $5\times 10^{-13}$&$1.46 \times 10^{-7}$ & $1.22\times 10^{-7}$ &$1.41\times 10^{-7}$ \\

SAX J$1808$-$3658$ &$\frac{40}{2\times 365}$& $2.5\times 10^{-14}$ &$3.20 \times 10^{-7}$  & $2.66 \times 10^{-7}$ &$3.08 \times 10^{-7}$\\

XTE J$1814$-$338$ &$\frac{50}{19 \times 365}$&$1.5\times 10^{-14}$  &$2.12\times 10^{-7}$ &  $1.76 \times 10^{-7}$&$2.04 \times 10^{-7}$\\
\tableline
\end{tabular}}
\tablecomments{Duty cycle $\Delta$, spin-up rate $\dot{\nu}_{su}$
\citep{Patruno:2012ab}, and upper limits on {\it r}-mode amplitudes from the
``spin-equilibrium'' condition $\alpha_{sp.eq} $, for three neutron star transients with
assumed masses of $1.4\, M_{\odot}$, $2.0\, M_{\odot}$ and $2.21\,
M_{\odot}$.}
\end{table}

\section{Constraining the {\it\lowercase{r}}-mode amplitude}

\subsection{Constraints from ``Spin Equilibrium''}
Here we compare two different methods for constraining the {\it r}-mode
amplitude, $\alpha$, from observations of LMXB NS
transients.  The first one, which gives larger values for $\alpha$, is
based on the spin equilibrium assumption
\citep{2000ApJ...536..915B,Ho:2011tt,2011ApJ...738L..14H,Watts:2008qw}
where we assume that in an outburst-quiescence cycle all the spin-up
torque due to accretion during the outburst is balanced by the {\it r}-mode
spin-down torque due to gravitational radiation in the whole
cycle. This is similar to the prescription considered by previous
authors \citep{2000ApJ...536..915B,Ho:2011tt}, but rather than using a
``fiducial'' torque estimated from the long-term average $\dot{M}$, we
can now use the observed spin-up rates and outburst properties to
directly constrain the torque.  Therefore we have
\begin{equation}
2\pi I \dot{\nu} \Delta= \frac{2 J_c}{\tau_G}\label{eq:spin-equilibrium}
\end{equation}
where $\dot{\nu}$ is the spin-up rate during outburst and
$\Delta=\frac{t_o}{t_r}$ is the ratio of the outburst duration, $t_o$,
to the recurrence time, $t_r$. We can estimate the left hand side of
Equation~(\ref{eq:spin-equilibrium}) from X-ray observations of LMXBs and
since the right hand side is a function of $\alpha$ through $J_c$ (see
Equation~(\ref{eq:J_c})) we can determine $\alpha$. In
Table \ref{tab:spin-equilibrium alpha} we give the results for $\alpha$
for three sources for which there are now estimates of the spin-up
rate due to accretion. The values of $\alpha$ are given for the three
different NS models considered in this work, and they are
all in the range of $\approx 1 - 3 \times 10^{-7}$.

\begin{table*}
\renewcommand{\arraystretch}{1.5}
\caption{Estimates of NS Temperatures and Luminosities\label{tab:temperature-luminosities}}
\scalebox{0.99}{
\begin{tabular}{crrrrrrrrrr}
\tableline\tableline
Source & $\nu_s$  (Hz) & Distance &$kT_{eff} $ &$T_{core} (K)$&$T_{core} (K)$&$L_{\gamma} (erg.s^{-1})$&$L_{\nu}(erg.s^{-1})$&$L_{\gamma}(erg.s^{-1})$&$L_{\nu}(erg.s^{-1})$\\
&&(kpc)&(eV)&$(1.4\, M_{\odot})$&$(2.21\, M_{\odot})$&$(1.4\, M_{\odot})$&$(1.4\, M_{\odot})$&$(2.21\, M_{\odot})$&$(2.21\, M_{\odot})$\\
\tableline
$4$U $1608$-$522$ & $620$&$4.1\pm 0.4$&$170$& $1.25\times 10^{8}$&$1.34 \times 10^8$ &$1.21\times 10^{34}$&$2.16\times 10^{32}$&$3.03\times 10^{34}$&$2.20\times 10^{39}$\\
IGR J$00291$+$5934$ & $599$&$5\pm 1$&$71$& $2.96\times 10^{7}$&$3.18 \times 10^7$ &$3.65\times 10^{32}$&$2.06\times 10^{27}$&$9.27\times 10^{32}$&$3.87\times 10^{35}$\\
MXB $1659$-$29$&$556$&$11.5\pm 1.5$&$55$&$1.96 \times 10^{7}$&$2.07 \times 10^{7}$ & $1.35\times 10^{32}$&$7.64\times 10^{25}$&$3.29\times 10^{32}$&$2.96\times 10^{34}$\\
Aql X-1& $550$&$4.55\pm 1.35$&$94$& $4.70\times 10^{7}$ & $5.06 \times 10^7$&$1.12\times 10^{33}$&$8.40\times 10^{28}$&$2.87\times 10^{33}$&$6.35\times 10^{36}$\\
KS $1731$-$260$&$524$&$7.2\pm 1.0$& $70$&$2.89 \times 10^{7}$&$3.12 \times 10^{7}$ &$3.44\times 10^{32}$&$1.70\times 10^{27}$&$8.88\times 10^{32}$&$3.47\times 10^{35}$\\
XTE J$1751$-$305$&$435$&$9 \pm 3$&$<71$& $2.96\times 10^{7}$&$3.18\times 10^7 $ &$3.65\times 10^{32}$&$2.06\times 10^{27}$&$9.27\times 10^{32}$&$3.87\times 10^{35}$\\
SAX J$1808$-$3658$ & $401$&$3.5\pm 0.1$&$<30$& $7.23\times 10^{6}$& $7.69 \times 10^6$ &$1.21\times 10^{31}$&$2.63\times 10^{22}$&$2.99\times 10^{31}$&$7.78\times 10^{31}$\\
XTE J$1814$-$338$&$314$&$6.7 \pm 2.9$&$<69$&$2.82\times 10^{7}$&$3.01 \times 10^7 $&$3.24\times 10^{32}$&$1.39\times 10^{27}$&$8.12\times 10^{32}$&$2.78\times 10^{35}$\\
NGC $6440$&$205$&$8.5 \pm 0.4$&$87$&$4.11 \times 10^{7}$&$4.46 \times 10^{7}$ & $8.10\times 10^{32}$&$2.88\times 10^{28}$&$2.11\times 10^{33}$&$2.97\times 10^{36}$\\
XTE J$1807$-$294$&$191$&$8.35 \pm 3.65$&$<51$&$1.72\times 10^{7}$&$1.83\times 10^7 $ &$9.84\times 10^{31}$&$2.71\times 10^{25}$&$2.46\times 10^{32}$&$1.44\times 10^{34}$\\
XTE J$0929$-$314$&$185$&$7.8 \pm 4.2$&$<50$&$1.66 \times 10^{7}$& $1.79 \times 10^7 $ &$9.06\times 10^{31}$&$2.06\times 10^{25}$&$2.31\times 10^{32}$&$1.23\times 10^{34}$\\
\tableline
\end{tabular}}
\tablecomments{Spin frequency, distance
to the source \citep{Watts:2008qw}, effective temperature at the
surface of the star \citep{Heinke:2006ie,Heinke:2008vj,Tomsick:2004pf},
core temperature, photon luminosity at the surface of the star and
neutrino luminosity for both $1.4\, M_{\odot}$ and $2.21\, M_{\odot}$
NSs. Note that $kT_{eff} $ given in this table is for a $1.4\,
M_{\odot}$ NS with a radius of $10$ km but in computing the core
temperatures and luminosities for different NS models the
appropriate redshifts have been used.  We note that for the sake of
brevity the core temperatures for the $2\, M_{\odot}$ models are not
included in the table; however, their values are all $\approx 5\%$ 
less than those for the $2.21\, M_{\odot}$ model.}
\end{table*}

At these amplitudes the inferred {\it r}-mode spin-down rate would be
competitive with the magnetic dipole spin-down rate which almost
certainly exists in these LMXBs, and which is quite likely the
dominant spin-down mechanism. Moreover, the amplitudes are also
comparable to those deduced assuming {\it r}-mode spin equilibrium with the
``fiducial'' accretion torques estimated by
\citet{2000ApJ...536..915B}. Those authors also demonstrated that at
such amplitudes some of these objects should have significantly higher
quiescent luminosities due to {\it r}-mode reheating than observed.  These
results suggest that the {\it r}-mode torque does not balance the
accretion-driven spin-up torque and that {\it r}-mode amplitude estimates
based on the ``spin equilibrium'' assumption will overestimate the
true amplitude.



\subsection{Constraints from ``Thermal Equilibrium''}

Here, we use the same thermal equilibrium argument outlined by Brown
\& Ushomirsky, but rather than estimating the quiescent luminosity
using the {\it r}-mode amplitude deduced from ``spin equilibrium,'' we use
observations of the quiescent luminosity of LMXBs to directly
constrain the amplitude of the {\it r}-mode.  This works because in a
steady-state, gravitational radiation pumps energy into the {\it r}-mode at
a rate given by, $W_d = (1/3) \Omega \dot J_c = -2E_c/ \tau_G$. This
expression has the familiar relationship for a power dissipated by an
applied torque, and in this case it is simply the {\it r}-mode torque due to
gravitational radiation. In a thermal steady-state all of this energy
must be dissipated in the star. Some fraction of this heat will be
lost from the star due to neutrino emission and the rest will be
radiated at the surface. It should be mentioned that the thermal
steady-state is not an assumption, but a rigorous result when the mode
is saturated, and in particular it is independent of the cooling
mechanism \citep{Alford:2012yn}. We further assume that all of the
energy emitted from the star during quiescence is due
to the {\it r}-mode dissipation inside the star. This is equivalent to
setting $H=0$ in Equation (12c). The resulting {\it r}-mode amplitude limits are
upper bounds in the sense that the observed luminosity reflects the
contribution from {\it r}-mode heating as well as any additional sources of
heat that are present, such as for example due to accretion and the
nuclear processing of accreted material, so-called deep crustal
heating \citep{2003A&A...404L..33H, 2007ApJ...662.1188G}.  If any
such sources of heat are present then the actual {\it r}-mode amplitude will
be less than the upper bounds given here. For the sources that we
study in this work, since we know the values of surface temperatures
and quiescent luminosities from observations, we can estimate the core
temperature and therefore determine the neutrino luminosities to
estimate the total amount of heat deposited in the core of these
systems by gravitational radiation.

To compute the core temperatures we use equation A$8$ in
\citet{Potekhin:1997mn} which relates the effective surface temperature
of the star $T_{eff}$ to the internal temperature $T_b$, which is the
temperature at a fiducial boundary at $\rho_b=10^{10}$ g cm$^{-3}$ for
a fully accreted envelope and is valid at not too high temperatures
$(T_b \leq 10^8 K)$,
\begin{equation}
(\frac{T_{eff}}{10^6 K})^4=(\frac{g}{10^{14} cm s ^{-2}})(18.1\frac{T_b}{10^9 K})^{2.42}\label{eq:T_b potekhin}
\end{equation}
where $g=GM/(R^2 \sqrt{1-r_g/R})$ is the surface gravity and
$r_g=2GM/c^2$.  Here we assume that the NS's core is
isothermal and since the thermal conductivity of the crust is high
\citep{Brown:2009kw} we have $T_{core}=T_b$ to good approximation.  If
we instead use equation A$9$ in \citet{Potekhin:1997mn} which gives
$T_b$ for a partially accreted envelope, and a column depth of
$\frac{P}{g}=10^9$ g cm$^{-2}$ \citep{2012MNRAS.424...93H} we get core
temperatures slightly higher than those derived from Equation (\ref{eq:T_b
potekhin}). The right hand side of the error bars on the temperatures
in Figure~\ref{fig:Omega-T} shows this difference. It is really only
relevant for a single source, 4U 1608-522, but even in this case it is
less than a $50\%$ increase, and the difference is always small enough
that it doesn't qualitatively change our results in the remainder of
the paper.  To compute $T_b$ we have used the effective surface
temperatures, $T_{eff}$, given by
\citet{Heinke:2006ie,Heinke:2008vj} and \citet{Tomsick:2004pf}\footnote{It should
be emphasized that the temperatures given in Table 2 of
\citet{Heinke:2006ie,Heinke:2008vj} are effective temperatures at the
surface of the star and not redshifted surface temperatures seen by an
observer at infinity. We note that they have been incorrectly assumed
to be redshifted temperatures in \citet{Degenaar:2012tw} and
\citet{2012MNRAS.424...93H}.  }. Note that those temperatures are
computed for a $1.4\, M_{\odot}$ NS with a $10$ km radius and we have
used the appropriate redshifts to compute $T_{core}$ for our neutron
star models.

Having surface and core temperatures for these sources we can use
Equations~(\ref{eq:neutrino-luminosity}) and (\ref{eq:thermal-luminosity}) to
evaluate their neutrino and thermal luminosities for different stellar
models. The values of the core temperatures, as well as the photon and
neutrino luminosities for different sources for the $1.4\, M_{\odot}$
and $2.21\, M_{\odot}$ NS models are computed and given in
Table~\ref{tab:temperature-luminosities}. Note that in the case of
$1.4\, M_{\odot}$ NSs the neutrino luminosity is only due to the
modified Urca reactions, but for the $2.21\, M_{\odot}$ NS it is due
to both direct and modified Urca reactions. By comparing the neutrino
and thermal luminosities in Table \ref{tab:temperature-luminosities}
one can see that at temperatures relevant for the LMXBs the standard
neutrino cooling from modified Urca reactions is negligible compared
to the photon emission from the surface of the star. However, if the
star is massive enough to enable direct Urca reactions in the core
(such as for our $2.21\, M_{\odot}$ NS model) then the
neutrino emission will dominate the cooling process for surface
temperatures higher than about $34$ eV \citep{Brown:1998qv}.

The thermal equilibrium condition can be written as $W_d =
L_{\nu}+L_{\gamma}$, where reheating due to {\it r}-mode dissipation is
given by $W_d=\frac{-2E_c}{\tau_{GR}}$ and is a function of {\it r}-mode
amplitude, $\alpha$. Therefore, $\alpha$ can be written in terms of
luminosities as
\begin{equation}
\alpha=\frac{5\times 3^4}{2^8 \tilde{J} M R^3 \Omega^4}(\frac{L_{\gamma}+L_{\nu}}{2 \pi G})^{1/2}\label{eq:alpha_thermal}
\end{equation}
where $L_{\gamma} = 4\pi R^2 \sigma T_{eff}^4$ is the thermal photon
luminosity at the surface of the star. Here, $R$ and $T_{eff}$ are the
stellar radius and surface temperature, respectively, and the neutrino
luminosity is given by
\begin{equation}
L_{\nu}=\frac{4 \pi R^3\Lambda_{QCD}^3 \tilde{L}_{DU}}{\Lambda_{EW}^4}T^6+\frac{4 \pi R^3 \Lambda_{QCD} \tilde{L}_{MU}}{\Lambda_{EW}^4}T^8
\end{equation}
where $T$ is the core temperature, $R_{DU}$ is the radius of the core
where direct Urca neutrino emission is allowed and $\tilde{L}$ is a
dimensionless parameter given in
Table~\ref{tab:viscosity-parameters}. The thermal equilibrium
condition for an NS with standard neutrino cooling ($1.4\, M_{\odot}$
and $2.0\, M_{\odot}$ NSs in this study) can be approximated as $W_d
\simeq L_{\gamma}$, since the neutrino cooling in this case is
negligible compared to the surface photon luminosity.

\begin{table*}
\renewcommand{\arraystretch}{1.5}
\caption{Upper Bounds on {\it\lowercase{r}}-mode Amplitudes and NS Spin-down Rates\label{tab:alpha} }
\begin{center}
\begin{tabular}{crrrrrrrr}
\tableline\tableline
Source & $\alpha_{th.eq}$ & $\alpha_{th.eq}$ & $\alpha_{th.eq} $ & $\dot{\nu}$ (Hz s$^{-1}$) &$\dot{\nu}$ (Hz s$^{-1}$) &$\dot{\nu}$ (Hz s$^{-1}$) &$\dot{\nu}_{sd}$ (Hz s$^{-1}$) \tabularnewline
&$(1.4\, M_{\odot})$&$(2.0\, M_{\odot})$&$(2.21\, M_{\odot})$ &$(1.4\, M_{\odot})$&$(2.0\, M_{\odot})$&$(2.21\, M_{\odot})$&observation \\
\tableline
$4$U $1608$-$522$ &$7.15\times 10^{-8}$ &$6.60\times 10^{-8}$ &$2.61\times 10^{-5}$ &-$1.44\times 10^{-15}$ &-$1.78\times 10^{-15}$ &-$2.08\times 10^{-10}$ &\\
IGR J$00291$+$5934$ & $1.41\times 10^{-8}$ &$1.32\times 10^{-8}$ &$3.99\times 10^{-7}$ & -$4.42\times 10^{-17}$ &-$5.59\times 10^{-17}$ &-$3.82\times 10^{-14}$ &-$3\times 10^{-15}$ \\
MXB $1659$-$29$&$1.16\times 10^{-8}$ &$1.07\times 10^{-8}$ &$1.49\times 10^{-7}$ & -$1.78\times 10^{-17}$ &-$2.18\times 10^{-17}$ &-$3.16\times 10^{-15}$ &\\
Aql X-1&$3.49\times 10^{-8}$ &$3.27\times 10^{-8}$ &$2.26\times 10^{-6}$ & -$1.49\times 10^{-16}$ &-$1.89\times 10^{-16}$ &-$6.74\times 10^{-13}$ &\\
KS $1731$-$260$&$2.35\times 10^{-8}$ &$2.20\times 10^{-8}$ &$6.44\times 10^{-7}$ & -$4.81\times 10^{-17}$&-$6.09\times 10^{-17}$&-$3.90\times 10^{-14}$&\\
XTE J$1751$-$305$&$5.09\times 10^{-8}$ &$4.76\times 10^{-8}$ &$1.44\times 10^{-6}$ & -$6.13\times 10^{-17}$ &-$7.74\times 10^{-17}$ &-$5.29\times 10^{-14}$ &-$5.5\times 10^{-15}$\\
SAX J$1808$-$3658$ &$1.28\times 10^{-8}$ &$1.19\times 10^{-8}$ &$3.30\times 10^{-8}$ & -$2.19\times 10^{-18}$ &-$2.74\times 10^{-18}$ &-$1.57\times 10^{-17}$ &-$5.5\times 10^{-16}$\\
XTE J$1814$-$338$ &$1.76\times 10^{-7}$ &$1.67\times 10^{-7}$ &$4.49\times 10^{-6}$ & -$7.49\times 10^{-17}$ &-$9.73\times 10^{-17}$ &-$5.26\times 10^{-14}$ &\\
NGC $6440$&$1.54\times 10^{-6}$ &$1.45\times 10^{-6}$ &$8.03\times 10^{-5}$ & -$2.90\times 10^{-16}$ &-$3.71\times 10^{-16}$ &-$8.50\times 10^{-13}$ &\\
\tableline
\end{tabular}
\end{center}
\tablecomments{Upper bounds on the {\it r}-mode amplitude from the ``thermal
equilibrium'' condition that are consistent with quiescent luminosity
data are given for different neutron star models. The
gravitational radiation induced spin-down rates due to unstable
{\it r}-modes as well as the observed spin-down rate for some of the sources
are also given \citep{Patruno:2010qz,Patruno:2012ab,Patruno:2011gu}.}
\end{table*}

\begin{figure*}
\begin{center}
\includegraphics[scale=0.7]{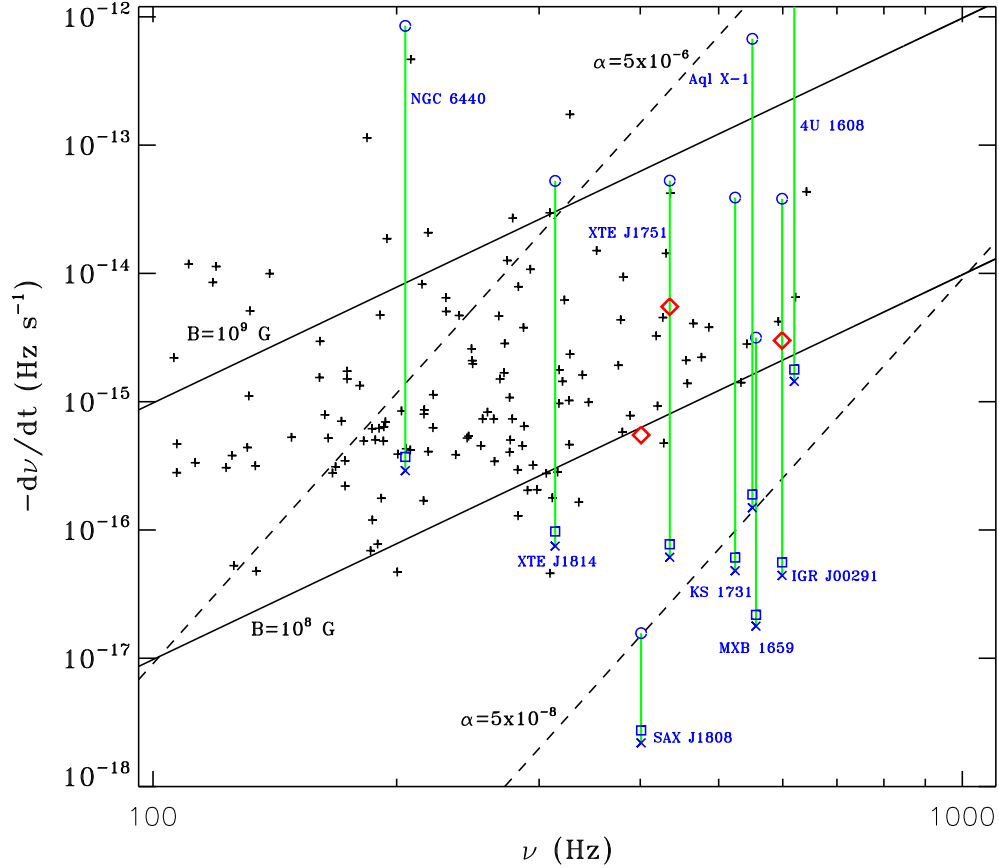}%
\end{center}
\caption{\label{fig:nu-ndot} Limits on the spin-down rates due to an
{\it r}-mode torque for nine LMXB systems and a range of NS masses
are shown in the $\dot{\nu}$ vs. $\nu$ plane. The $\dot\nu$ limits
for $1.4$, $2.0$, and $2.21\, M_{\odot}$ NS models obtained from the
{\it r}-mode amplitude limits derived from observations of quiescent
luminosities and temperatures (see the discussion in Section 3) are
marked by the $\times$, square, and circle symbols, respectively. The
vertical green lines connecting the symbols show the full range of
$\dot{\nu}$ for each source (labeled).  Also shown are two pairs of
parallel lines representing magnetic-dipole (solid) and {\it r}-mode
(dashed) braking laws. Lines are drawn for two values of the magnetic
field, $10^8$ G (lower), and $10^9$ G (upper), as well as two values
of $\alpha$, $5\times 10^{-8}$ (lower) and $5\times 10^{-6}$
(upper). For systems with measured, quiescent spin-down rates, these
values are marked with the red diamond symbols. For additional
context, millisecond pulsars from the ATNF pulsar database
\citep{Manchester:2004bp} are shown and denoted by the black $+$
symbols.\\
(A color version of this figure is available in the online journal.)}
\end{figure*}

As can be seen in Figure~\ref{fig:Omega-T}, out of the $11$ sources
considered in this paper all but $2$ of them are likely to have
unstable {\it r}-modes, meaning that they are above the {\it r}-mode instability
curve. The two most slowly rotating sources, XTE J$1807$ and XTE
J$0929$, are outside the instability region for our NS
models, which means they likely can no longer spin-down due to
gravitational radiation from an {\it r}-mode\footnote{It has been shown by
\citet{Alford:2010fd} that the boundary of the instability region is
insensitive to the quantitative details of the microscopic
interactions that induce viscous damping in a given phase of dense
matter.}. Therefore, we only evaluate the upper bounds on the {\it r}-mode
amplitude for those nine sources within the instability window. Using
Equation~(\ref{eq:alpha_thermal}), we have evaluated $\alpha$ for all of
those sources using the three different NS models considered in this
work. The values of $\alpha$ are given in Table~\ref{tab:alpha}. As
can be seen for the $1.4\, M_{\odot}$ and $2.0\, M_{\odot}$ NSs, where
there is no enhanced neutrino emission, the values of $\alpha$ range
from $1.07\times 10^{-8}$ to $1.54\times 10^{-6}$, where NGC
$6440$--which has the lowest spin frequency--has the highest {\it r}-mode
amplitude, as expected. In the case of $2.21\, M_{\odot}$ NSs the
upper bounds on $\alpha$ are larger since due to the direct Urca
neutrino emission $W_d$ can be larger, and $\alpha$ in this case
ranges from $3.30\times 10^{-8}$ to $8.03\times 10^{-5}$. $4$U
$1608$-$522$ has the highest temperature among the sources considered
which implies a very large neutrino luminosity for the $2.21\,
M_{\odot}$ NS model and as a result a large {\it r}-mode amplitude. The
large value of the {\it r}-mode amplitude in this source may be ruled out by measurement of its spin-down rate, as
is explained further in the next section. The high temperature
of $4$U $1608$-$522$ could be explained if it has a lower mass, but it
should also be noted that this system has been accreting for a long
time and both its high quiescent luminosity and surface temperature
may be due to the long term accretion and our assumption in ascribing
all the quiescent luminosity to the heat that comes from inside of the
star may not be a good estimate for this system, but here we are
only interested in obtaining upper limits on {\it r}-mode amplitude.

\subsection{{\it r}-mode Spin-down}

To see whether or not these results are plausible and what fraction of
the quiescent spin-down of these sources can be due to gravitational
radiation from {\it r}-mode oscillations, we use our results for $\alpha$
from the ``thermal equilibrium'' condition and insert them into the
right hand side of Equation~(\ref{eq:evolution-Omega}) to determine
$\dot{\nu}$ in the absence of accretion ($N_{acc} = 0$).  The derived
spin-down values for different NS models are given in
Table~\ref{tab:alpha}, and are shown graphically in
Figure~\ref{fig:nu-ndot} (vertical green lines). Comparing these results
with the observed spin-down rates, which exist for IGR J$00291$, XTE
$1751$-$305$ and SAX J$1808$ (red diamonds in Figure~\ref{fig:nu-ndot}),
we find that in the case of $1.4\, M_{\odot}$ ($\times$ symbols) and
$2.0\, M_{\odot}$ (square symbols) NSs, where there is no fast
neutrino cooling present in the star, the {\it r}-mode spin-down can only
provide about $1\%$ of the observed spin-down rate, which means that
other spin-down mechanisms such as magnetic-dipole radiation are
responsible for spinning down a $1.4\, M_{\odot}$ or $2.0\, M_{\odot}$
hadronic star with no fast cooling process. For the two AMXPs with
relatively ``slow'' spin frequencies, NGC 6440 and XTE J1814-338, we
find {\it r}-mode spin-down limits that are more competitive with observed
values.  Indeed, our limit for NGC 6440 is comparable to the measured,
quiescent spin-down rate for SAX J1808, and thus spin-down measurements
for this source would be particularly interesting.

On the other hand, the {\it r}-mode amplitudes we obtain for the $2.21\,
M_{\odot}$ (circle symbols) hadronic model with direct Urca neutrino
emission are only consistent with observations for SAX J1808 (and
perhaps MXB 1659), as the inferred spin-down rates are either less
than the observed rate for the source--in the case of SAX J1808--or
similar to the other observed rates, as for MXB 1659.  For the
remaining sources considered here the $2.21\, M_{\odot}$ limits are
likely not consistent with the observations since such large
amplitudes imply very large {\it r}-mode spin-down rates, and in the case of
IGR J$00291$ and XTE J$1814$ they are in fact larger than the observed
values. If the neutrino luminosity from these sources was indeed as
large as estimated with our $2.21\, M_{\odot}$ model, then in thermal
equilibrium there must be a heat source that can supply it. Since the
spin-down measurements for these sources indicate that {\it r}-mode heating
(for this model) would be insufficient, several possibilities
remain. First, there could be some additional source of heat other
than {\it r}-mode dissipation that supplies the needed energy.  However, we
note that it would need to supply a substantial luminosity, as the
direct Urca neutrino emission for this model outshines the photon
luminosity by more than an order of magnitude, and we are not aware of
any simple mechanisms that could provide the required
luminosity. Second, the actual mass of these systems could be less
than that of the model in question ($2.21\, M_{\odot}$). Indeed, if it
could be demonstrated that {\it r}-mode dissipation were the only mechanism
that could produce such a large luminosity then an upper limit on the
mass would follow, and the limit would be the mass for which the
neutrino plus photon luminosity matched the {\it r}-mode heating produced
when the amplitude is large enough to produce a spin-down rate equal
to the observed quiescent rate or our theoretical value for a
high mass NS model, whichever is smaller.  This would be a
conservative limit in the sense that it is likely that the {\it r}-mode
torque does not account for all of the observed spin-down.  Finally,
our model assumptions, for example, the EOS and core composition,
could be incorrect, with one possibility being the existence of exotic
matter, such as kaon or pion condensates, or quark matter in the core
which have smaller neutrino emissivities than nucleon direct Urca
processes, or if the pairing gaps for $^3P_2$ neutrons and
$^1S_0$ protons were larger than current theoretical values (this will
be explained in more detail in the next paragraph).  Interestingly,
if the masses of these systems were known then one of the
possibilities outlined above is precluded and then the observationally
derived {\it r}-mode limits become sensitive to properties of the core,
either the presence of exotic matter or perhaps additional heating
physics.  In this sense further spin-down measurements, and where
possible, mass constraints could provide interesting new insights on
the physics of dense NS matter.

Our theoretical treatment of neutrino emission processes in this
study, namely modified and direct Urca, spans the plausible range
between ``slow'' and ``fast'' neutrino cooling processes. Here we have
considered NS models made of non-superfluid hadronic matter with the
APR EOS, but more realistically it is likely that
neutrons and protons will be in a superfluid phase inside
NSs. Therefore, a natural question would be whether or not our conclusions
will still hold in the presence of superfluidity. Considering
the presence of superfluidity in these sources, assuming that they are
still inside the unstable region for {\it r}-modes, could have two possible
effects on their cooling. The first one is neutrino emission due to
Cooper pair breaking and formation (PBF) just below the superfluid
critical temperature \citep{ Page:2004fy} and the second one is the
suppression of direct and modified Urca neutrino cooling at
temperatures below the critical temperature. Here we explain why our
qualitative results, and our argument about setting upper bounds on
the masses of these sources, are not changed by considering the effect
of superfluidity on the neutrino cooling of these sources.  In our low
mass NS models ($M<2~M_{\odot}$) where there is no fast cooling
mechanism in the core, the thermal luminosity is much larger than the
modified Urca neutrino luminosity, in fact by more than five orders of
magnitude in all of the sources considered here except 4U
1608. Therefore, even if the temperature of these sources were just
below the critical temperature of superfluidity, neutrino emission due
to PBF, which is only about 10 times stronger than modified Urca
neutrino emission, would still be much smaller than the thermal
emission. Therefore, suppression of neutrino emission, or neutrino
emission due to PBF is not important for our low mass NS models. What
about high mass NSs where fast cooling can happen in the core? Since
direct Urca neutrino emission is much stronger than neutrino emission
due to PBF, if direct Urca is not suppressed by superfluidity, it will
be the dominant cooling mechanism. Now the question is whether or not neutron
and proton pairing can suppress direct Urca neutrino emission in the
core. Current theoretical results for the pairing gaps in
$^3P_2$ neutron superfluid \citep{Schwenk:2003bc, Dong:2013sqa} and
$^1S_0$ proton superconducting phases (see for example
\citet{Page:2004fy,Yakovlev:2004iq}), which are relevant in the core
of NSs, suggest that both of these
pairing gaps are vanishingly small in the inner core, where the direct Urca process can operate. Therefore, superfluidity is unlikely to
suppress fast cooling processes in the core of NSs and thus neglecting
the effect of superfluidity does not change our qualitative results,
assuming that {\it r}-modes are still unstable in the presence of
superfluidity in these sources.

We also note that the density profile (density versus radius) for an NS
made of hadronic matter with the APR EOS is almost flat at the center
of the star, which means that as the mass of the star increases above
$2 M_{\odot}$ (above which direct Urca processes can operate in the
core of an NS made of hadronic matter with the APR EOS), there will be
a sizable region in the core where direct Urca processes may happen,
which can make it easier to obtain an upper limit on the mass of these
sources using spin-down measurements.

With typical values of a few $\times 10^{-8}$, our derived amplitude
upper limits suggest that for many LMXB NSs the {\it r}-modes are
likely not excited to sufficient amplitudes to substantially affect
their spin evolution.  This begs the question of whether or not
unstable, steady-state {\it r}-modes actually exist in these NSs.
One possibility is that additional damping mechanisms, such as those perhaps
associated with crust effects, such as the viscous friction
at the crust-core boundary due to the coupling between core {\it r}-modes
and crustal torsional modes \citep{Levin:2000vq}, superfluid mutual
friction \citep{Ho:2011tt} or the existence of exotic matter in the
core of NSs \citep{Alford:2010fd,Schwenzer:2012ga}, are at work and
modify the instability window so as to render these NSs
stable to {\it r}-mode excitation.  An interesting related question is
whether the existence of {\it r}-modes at the amplitudes estimated here can
be inferred directly from observations. Figure~\ref{fig:nu-ndot} shows
both {\it r}-mode (dashed parallel lines) and magnetic-dipole (solid
parallel lines) spin-down laws.  The {\it r}-mode spin-down braking index,
$n=7$, is steeper compared to that for magnetic spin-down ($n=3$),
thus at high enough spin frequencies (well above a kHz) one might
expect that the {\it r}-mode torque would eventually become competitive with
or dominate the magnetic dipole torque.  However, as of yet there are
no known NSs spinning fast enough for this effect to become
dominant, and depending on the EOS the mass-shedding limit might be
reached before the {\it r}-mode torque becomes competitive with the magnetic
torque.

As discussed above, quiescent spin-down measurements have been
typically attributed to the magnetic torque.  For a number of sources
considered here our {\it r}-mode amplitude limits support this
presumption. Any spin-down contribution from an {\it r}-mode torque would be
more easily identifiable if the magnetic field strengths of these
NSs were constrained independently from the magnetic
spin-down estimate.  Moreover, identifying an {\it r}-mode spin-down would,
in principle, be simpler for those NSs with the lowest
magnetic torques, and thus field strengths.  At present, the lowest
inferred dipolar magnetic field strengths are $\approx 6
\times 10^{7}$ G. At this level the magnetic spin-down is of the order
of $\dot{\nu} \approx 5 \times 10^{-17}$ Hz s$^{-1}$, which is comparable
to our derived {\it r}-mode spin-down limits for a number of sources
considered here, assuming that their NSs are $< 2\, M_{\odot}$. In
this regard, quiescent spin-down measurements for more of the sources
considered here, in particular the AMXPs XTE J1814 and NGC
6440, would be extremely valuable.

\subsection{Gravitational Wave Amplitudes}

{\it r}-modes in NSs are one of the possible mechanisms for gravitational
wave (GW) emission and they can be observationally interesting in
newborn NSs and perhaps accreting NSs in
LMXBs \citep{Owen:2010ng}. Continuous GW emission from {\it r}-modes is
dominated by $l=m=2$ current quadrupole emission
\citep{Lindblom:1998wf}.  The gravitational wave amplitude $h_0$
(strain tensor amplitude) is related to the {\it r}-mode amplitude $\alpha$
by the following equation \citep{Owen:2010ng,Owen:2009tj}
\begin{equation}
h_0=\sqrt{\frac{8\pi}{5}}\frac{G}{c^5}\frac{1}{r}\alpha\omega_r^3MR^3\tilde{J},
\end{equation}

where $r$ is the distance to the source, $M$ and $R$ are the mass and
radius of the NS, $\tilde{J}$ is the dimensionless parameter defined
by Equation~(\ref{eq:J-tilde}) and $\omega_r$ is the frequency of the {\it r}-mode,
which is related to the angular spin frequency of the star $\Omega$
(for the $m=l=2$ {\it r}-mode) by the following equation
\begin{equation}
\omega_r\approx \frac{4}{3}\Omega \; .
\end{equation}
Using the upper limits on the {\it r}-mode amplitude of NSs in LMXBs derived
above, we can obtain upper limits on the amplitude of the GWs emitted
from these sources due to unstable {\it r}-modes. For the sources considered
in this work, upper limits on the GW strain amplitude $h_0$ for the
$1.4\, M_{\odot}$ and $2.0\, M_{\odot}$ NS models are in the range of
$1.8\times 10^{-29}$ to $4.9\times 10^{-28}$ which is below the
anticipated detectability threshold of Advanced LIGO
\citep{Collaboration:2009rfa,Watts:2008qw}. In the case of the $2.21\,
M_{\odot}$ NS model, since the {\it r}-mode amplitudes are larger than those
for the low mass NSs, we get larger values of $h_0$, but even in this
case for most of the sources $h_0$ is still below the detectability
threshold of Advanced LIGO. The highest values of $h_0$ for a massive
star are obtained for NGC $6440$ and Aql X-$1$ with an amplitude of
the order of $8.5 \times 10^{-27}$ and $4$U $1608$-$522$ with an
amplitude of $1.59 \times 10^{-25}$. However, it should be noted that
the large {\it r}-mode amplitudes in these sources, which would cause very
large spin-down rates, may eventually be ruled out by future spin-down
measurements. In this context it is important to restore the X-ray
timing capability that was lost when {\it RXTE} was decommissioned in
January 2012. Missions planned or currently in development which could
provide such a capability include India's {\it ASTROSAT}
\citep{2012hcxa.confE..95S}, ESA's {\it Large Observatory for X-ray Timing}
\citep{2012SPIE.8443E..2DF}, and NASA's {\it Neutron Star Interior
Composition Explorer} \citep{2012SPIE.8443E..13G}.

The upper limits on GW amplitudes discussed here are related to the GW
emission due to unstable {\it r}-modes but our results do not exclude the
possibility of having larger GW amplitudes in LMXBs from other GW
emission mechanisms such as NS mountains \citep{Ushomirsky:2000ax,
Haskell:2006sv}.  It is worth mentioning that indirect upper limits on
GW amplitude can be obtained for sources with observed spin-down
rates, $\dot{\nu}$, by assuming that all of the observed spin-down is
due to GW emission \citep{Owen:2010ng},
\begin{equation}
h_0^{sd}=\frac{1}{r}\sqrt{\frac{45G I \dot{P}}{8c^3P}}
\end{equation}
where $P=\frac{2\pi}{\Omega}$ is the observed pulse period and
$|\frac{\dot{P}}{P}|=|\frac{\dot{\nu}}{\nu}|$.  Using this equation
for the three sources with measured $\dot{\nu}$ and different NS
models (i.e. different masses and radii) we obtain $h_0^{sd}$ values that
range from $ 4.14 \times 10^{-28}$ to $ 6.53 \times 10^{-28}$.

\section{Conclusions}
In this paper we have presented upper limits on {\it r}-mode amplitudes in
LMXB NSs using their observed quiescent luminosities,
temperatures and spin-down rates. We calculated results for NS models
constructed with the APR EOS (normal hadronic matter) with masses of
1.4, 2 and 2.21\, $M_{\odot}$, where our highest mass model (2.21
$M_{\odot}$) can support enhanced, direct Urca neutrino emission in
the core. We have used two different methods to calculate {\it r}-mode
amplitudes. The first is based on the assumption that in an
outburst-quiescence cycle all the spin-up torque due to accretion
during the outburst is balanced by the {\it r}-mode spin-down torque due to
gravitational radiation. This method gives amplitudes in the range of
$\approx 1 - 3 \times 10^{-7}$ for the sources with measured spin-up
rates. Since in reality there are other sources of spin-down such as
magnetic-dipole radiation that may be the dominant spin-down source,
we use another method for computing the {\it r}-mode amplitude that does not
ascribe all of the spin-down of the star to gravitational radiation
and therefore gives tighter bounds on the amplitudes.  This second
method is based on the assumption that in a thermal steady-state some
fraction of the heat that is generated in the star due to {\it r}-mode
dissipation will be lost from the star by neutrino emission and the
rest will be radiated at the surface. This assumes that all of the
heat emitted from the surface of the star during quiescence is due to
the {\it r}-mode dissipation inside the star, and thus provides an upper
bound on the {\it r}-mode amplitude. We have computed core temperatures as
well as neutrino and thermal (photon) luminosities for LMXB sources
using measurements of the quiescent luminosities and surface
temperatures and showed that at temperatures relevant for LMXB neutron
stars, when there is no enhanced cooling mechanism, the cooling of the
star is dominated by photon emission from the surface (for $1.4$ and
$2.0\, M_{\odot}$ NS models), but in a massive star where direct Urca
neutrino emission is allowed, the cooling is dominated by neutrino
emission (for $T_{eff} \gtrsim 34$ eV).  For the lower mass NS models
(1.4 and 2 $M_{\odot}$) we find dimensionless {\it r}-mode amplitudes in the
range from about $1\times 10^{-8}$ to $1.5\times 10^{-6}$.
We note that none of the saturation mechanisms proposed so far can
saturate {\it r}-modes at these low amplitudes. Alternatively, the enhanced 
dissipation that would result from
the existence of exotic matter in NS interiors could shift the
instability window such that the LMXBs are perhaps stable to {\it r}-mode
excitation \citep{Alford:2010fd,Schwenzer:2012ga}.

For the AMXP sources with known quiescent spin-down rates these limits
suggest that $\lesssim 1\%$ of the observed rate can be due to an
unstable {\it r}-mode. Interestingly, the AMXP with the highest amplitude
limit, NGC 6440, could have an {\it r}-mode spin-down rate comparable to the
observed, quiescent rate for SAX J1808. Thus, quiescent spin-down
measurements for this source would be particularly interesting.
Having enhanced, direct Urca neutrino emission in the core of our
highest mass model ($2.21\, M_{\odot}$) means that the dissipated heat
in the star can be larger and therefore it can have higher {\it r}-mode
amplitudes. Indeed, the inferred {\it r}-mode spin-down rates at these
higher amplitudes are inconsistent with the observed spin-down rates
for some of the LMXB sources, such as IGR J00291 and XTE J1751-305. If
{\it r}-mode dissipation were the only mechanism available to produce this
high luminosity, then this could be used to put an upper limit on the
masses of these sources if they were made of hadronic matter.  Alternatively, it could be used to probe the
existence of exotic matter in them if the NS mass in these systems
were known. In this way, future spin-down and NS mass measurements for
the LMXB systems considered here, as well as for yet to be discovered
systems, could open a new window on dense matter in NS interiors. For
this as well as other reasons, we regard the re-establishment of a
sensitive X-ray timing capability as vital to the use of NSs as
natural laboratories for the study of dense matter.  Using the results
for {\it r}-mode amplitudes, the upper limits on gravitational wave
amplitude due to {\it r}-modes have been computed.  The upper limits on the
GW strain amplitude $h_0$ for the $1.4\, M_{\odot}$ and $2.0\,
M_{\odot}$ NS models are in the range of $1.8\times 10^{-29}$ to
$4.9\times 10^{-28}$ which is below the anticipated detectability
threshold of Advanced LIGO. In the case of the $2.21\, M_{\odot}$ NS
model, we obtain larger values for $h_0$, but even in this case for most
of the sources considered in this work, $h_0$ is still below the
detectability threshold of Advanced LIGO. Gravitational waves due to
other mechanisms such as NS mountains may have larger amplitudes in
these systems.

\acknowledgments
SM thanks Mark Alford, Andrew Cumming, Cole Miller and Kai Schwenzer
for helpful discussions. TS acknowledges NASA's support for high
energy astrophysics. SM acknowledges the support of the
U.S. Department of Energy through grant No. DEFG02- 93ER-40762.

\bibliography{LMXB-rmode}


\end{document}